\documentclass[a4paper]{article}  %a4paper>>> use this instead for A4 paper
\usepackage[top=0.85in,left=0.75in,footskip=0.75in]{geometry}

\usepackage[utf8]{inputenc} 
\usepackage[T1]{fontenc}
\usepackage{amsmath, amsfonts, amssymb}
\usepackage{graphicx, booktabs, multirow}
\usepackage[allcolors=black]{hyperref}
\usepackage[nolist]{acronym}
\usepackage{siunitx}
\usepackage{nicefrac}
\usepackage{float, enumerate, adjustbox}
\usepackage{stackengine}
\usepackage{setspace}

\usepackage{placeins} % for \Floatbarrier

\usepackage{colortbl}
\usepackage{bigdelim}
\newcolumntype{_}{>{\global\let\currentrowstyle\relax}}
\newcolumntype{^}{>{\currentrowstyle}}

\usepackage{diagbox}
\usepackage{comment}
\usepackage{lastpage,fancyhdr,graphicx}
\usepackage{setspace} 
\usepackage{float}
\begin{document} 
\begin{flushleft}
{\huge
\textbf\newline{Photoacoustic monitoring of blood oxygenation during neurosurgical interventions
}
}
\newline
% authors go here:
\\
Thomas Kirchner \textsuperscript{1,2,*}, 
Janek Gröhl \textsuperscript{1,3}, 
Niklas Holzwarth \textsuperscript{1,2}, 
Mildred A. Herrera \textsuperscript{4}, 
Tim Adler \textsuperscript{1,5}, 
Adrián Hernández-Aguilera \textsuperscript{4}, 
Edgar Santos \textsuperscript{4}, 
Lena Maier-Hein \textsuperscript{1,3}
\\
\bigskip
{\small
{\bf 1} Division of Computer Assisted Medical Interventions, German Cancer Research Center, Heidelberg, Germany.
\\
{\bf 2} Faculty of Physics and Astronomy, Heidelberg University, Heidelberg, Germany.
\\
{\bf 3} Medical Faculty, Heidelberg University, Heidelberg, Germany.
\\
{\bf 4} Department of Neurosurgery, Heidelberg University Hospital, Heidelberg, Germany.
\\
{\bf 5} Faculty of Mathematics and Computer Science, Heidelberg University, Heidelberg, Germany.
\\
\bigskip
*\, Please address your correspondence to Thomas Kirchner, e-mail: t.kirchner@dkfz-heidelberg.de
}
\end{flushleft}

\begin{abstract}
Multispectral \ac{pa} imaging is a prime modality to monitor hemodynamics and changes in \ac{so2}. Although \ac{so2} changes can be an indicator of brain activity both in normal and in pathological conditions, \ac{pa} imaging of the brain has mainly focused on small animal models with lissencephalic brains. Therefore, the purpose of this work was to investigate the usefulness of multispectral \ac{pa} imaging in assessing \ac{so2} in a gyrencephalic brain. To this end, we continuously imaged a porcine brain as part of an open neurosurgical intervention with a handheld \ac{pa} and \ac{us} imaging system \emph{in vivo}. Throughout the experiment, we varied respiratory oxygen and continuously measured arterial blood gases. The \ac{Sao2} values derived by the blood gas analyzer were used as a reference to compare the performance of linear spectral unmixing algorithms in this scenario. According to our experiment, \ac{pa} imaging can be used to monitor \ac{so2} in the porcine cerebral cortex. While linear spectral unmixing algorithms are well-suited for detecting changes in oxygenation, there are limits with respect to the accurate quantification of \ac{so2}, especially in depth. Overall, we conclude that multispectral \ac{pa} imaging can potentially be a valuable tool for change detection of \ac{so2} in the cerebral cortex of a gyrencephalic brain. The spectral unmixing algorithms investigated in this work will be made publicly available as part of the open-source software platform Medical Imaging Interaction Toolkit (MITK).
\end{abstract}

\acresetall

\section{Introduction}

A major application of \ac{pa} imaging is the monitoring of hemodynamics and changes in \ac{so2} \cite{wang2012photoacoustic}, which are indicators of brain activity \cite{fransson2005spontaneous, raichle2001default} or injury \cite{takano2007cortical}. In clinical practice, techniques for the monitoring of brain injury can vary widely with the specific application. While there are various techniques that can image hemodynamics, \ac{pa} imaging can potentially provide better functional information and higher resolution \cite{yao2014photoacoustic}. \ac{so2} is usually calculated via the estimation of abundances of oxygenated and deoxygenated hemoglobin chromophores \cite{jobsis1977noninvasive}. In multispectral \ac{pa} imaging, this concentration estimation is generally done by linear \ac{su} \cite{keshava2002spectral, li2008simultaneous, gerling2014real}, which involves solving a set of linear equations for the desired abundances of hemoglobin \cite{chance1988comparison}. While \ac{pa} imaging is widely used in small animal models with lissencephalic brains, larger and more complex brains remain challenging \cite{yao2014photoacoustic}. The purpose of this work was therefore to investigate the usefulness of multispectral \ac{pa} imaging in assessing \ac{so2} in a gyrencephalic brain. 

\section{Material and Methods}

To investigate the performance of \ac{so2} estimation by \ac{su} \emph{in vivo} in a neurosurgical setting, we performed \ac{pa} measurements during an open intervention on a porcine brain, which allowed us to image without the acoustic attenuation of the skull. In this setting, we were also able to take corresponding \ac{abg} \cite{mcfadden1968arterial} measurements and thus to compare the quantitative \ac{so2} estimation performance of numerical algorithms for \ac{su} \cite{tzoumas2014unmixing}, against a physiological \ac{Sao2} reference value.

\subsection*{Experimental setup} 
% hier in der Vergangenheit schreiben 
The \ac{pa} imaging modality used in this study was a custom hybrid \ac{pa} and \ac{us} system with a fast-tunable optical parametric oscillator (OPO) laser system (Phocus Mobile, Opotek, Carlsbad, USA) and a 7.5\,MHz linear \ac{us} transducer with 128 elements (L7-Xtech, Vermon, Tours, France), on a DiPhAs ultrasonic research platform (Fraunhofer IBMT, St. Ingbert, Germany) \cite{kirchner2016freehand}. The custom probe holder covered the transducer with a gold leaf to reduce transducer absorption artifacts. For optimal contrast to noise \cite{selection}, the \ac{pa} images were recorded at 760\,nm and 858\,nm, adding 798\,nm as an isosbestic reference. 

\begin{figure}[hbt!]
\centering
\includegraphics[width=\linewidth]{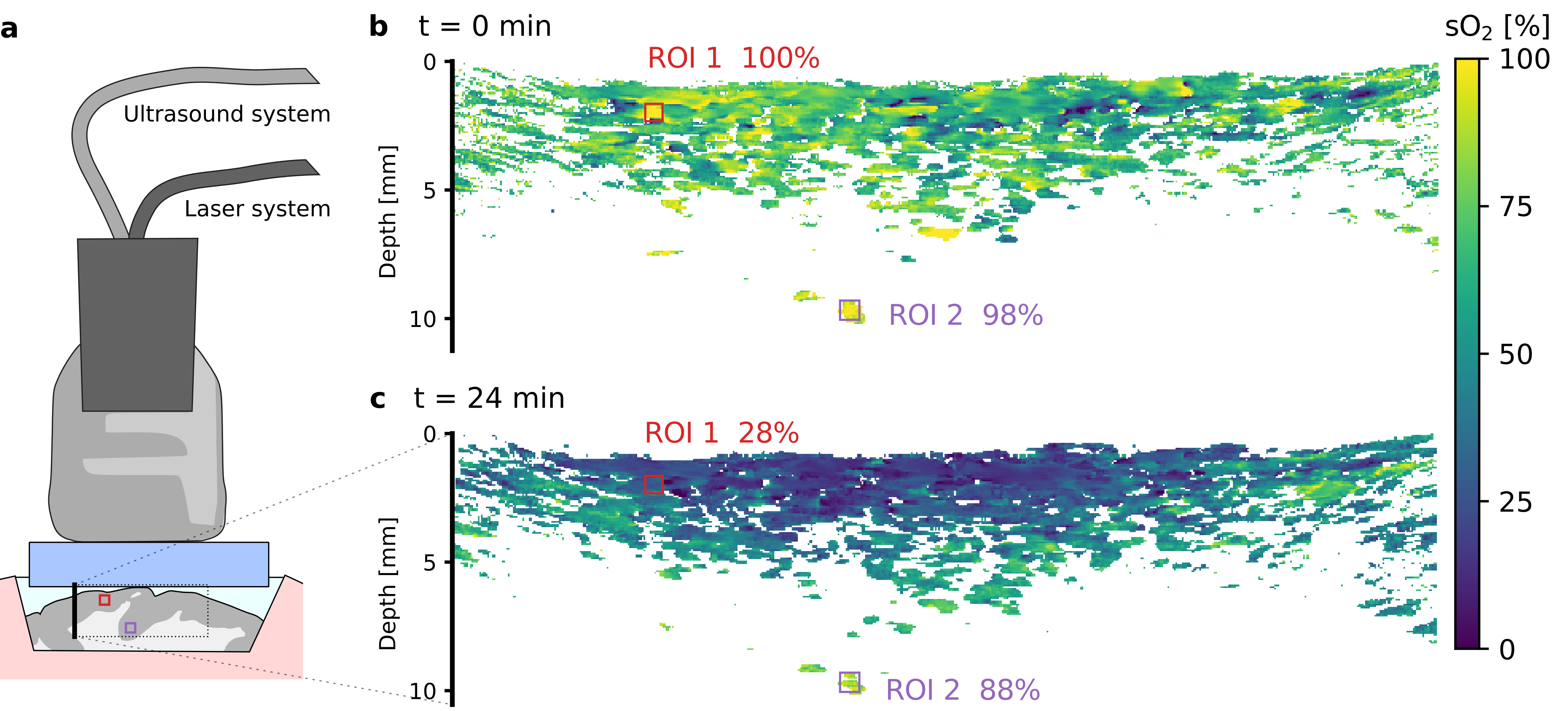}
\caption[]{\textbf{(a)} Experimental setup. The \acf{pa} probe was used to record sagittal images of a porcine brain. The field of view contains two regions of interest (\acp{roi}) corresponding to arteries near (ROI\,1 -- red) and far (ROI\,2 -- purple) from the brain surface. \textbf{(b)\,\&\,(c)} Example images of \acf{so2} in a brain slice. The \ac{roi}s are annotated with their corresponding median \ac{so2} value as determined with the Householder QR unmixing algorithm. $t$ is the time from start of the experiment. In both \ac{so2} images, values under a noise equivalent threshold of unmixed total hemoglobin were masked. In \textbf{(b)} the \acf{ro2} was 35\,\% and \acf{abg} analysis yielded an \acf{Sao2} of 100\,\%, while in \textbf{(c)} \ac{ro2} was 0\,\% with \ac{Sao2} = 26.2\,\%.}
\label{fig:example_static_oxy}
\end{figure}

A porcine brain was continuously imaged for 45\,min, as part of an open neurosurgical intervention with our hybrid \ac{pa} and \ac{us} probe. Our experiment was carried out following a craniotomy on a three month old female domestic pig. As illustrated in Figure \ref{fig:example_static_oxy}a, the probe was fixed over the left hemisphere of the brain to record a sagittal slice, using a gel pad as acoustic coupling. During imaging, the ventilation of the animal was varied and \ac{Sao2} and reference measurements with an \ac{abg} analyzer were taken. The ventilation changes and reference measurements are detailed in Figure \ref{fig:in_vivio} and Table \ref{tab:ROI}. In addition, \ac{Sao2} was monitored non-invasively with a pulse oximeter \cite{tremper1989pulse} placed on the left earlobe.

\subsection*{Image processing} 

Using the \ac{mitk} \cite{nolden2013medical} \ac{pa} image processing plugin \cite{kirchner2018signed}, the raw PA data was beamformed with delay and sum \cite{griffiths1982alternative,kim2016programmable} and von Hann apodization. The resulting images were motion corrected with the corresponding \ac{us} B-mode images. All input images were averaged over ten recordings per wavelength before linear \ac{su}, which was performed on the \ac{pa} images with five commonly used linear algorithms \cite{faires1994numerische}. To cover a wide range of algorithms, we selected a QR decomposition with Householder transformation \cite{goodall_1993}, a LU (with full pivoting) \cite{adomian_1988_review} and a singular value decomposition \cite{boardman_1989_inversion} all from the C++ \textit{Eigen} \cite{guennebaud2010eigen} library, as well as a weighted \cite{shimabukuro1991least} (based on QR decomposition) and a non-negative (using least angle regression \cite{efron2004least}) least square algorithm both from the C++ \textit{Vigra} \cite{vigra} library.
\acused{roi}

For quantitative validation of the \ac{su} algorithms, we selected two regions of interest (ROIs), for which we determined \ac{so2} values. We selected a surface \ac{roi} and a deep one to investigate the influence of fluence effects \cite{tzoumas2016eigenspectra} on \ac{su} \ac{so2} estimation. 
We assumed that both \acp{roi} contain arteries, as they had generally high PA signal and distinct characteristic pulsing in the \ac{us} and \ac{pa} image streams. The resulting \ac{so2} values for one \ac{roi} are the median of all pixels within that \ac{roi} that have a higher than noise equivalent total hemoglobin level.

\section{Results and Discussion}

According to our results in Table\,\ref{tab:ROI}, the different \ac{su} algorithms were similar in estimation performance, with the exception of the non-negative least square boundary effect. While other algorithms can yield  physiologically impossible \ac{so2} values (even above 100\,\%, and theoretically also below 0\,\%), the non-negativity constraint artificially prevents this. This was especially relevant for the evaluation of changes in \ac{roi}\,2. All other differences between the algorithms are within their respective standard deviations. The Householder QR algorithm performed the fastest. In the following, we therefore only present the \ac{su} results of the QR algorithm. Example slices are shown in Figure \ref{fig:example_static_oxy}b\&c with the marked \acp{roi} and their corresponding median \ac{so2} value.

\begin{figure}[h!]
\centering
\includegraphics[width=\linewidth]{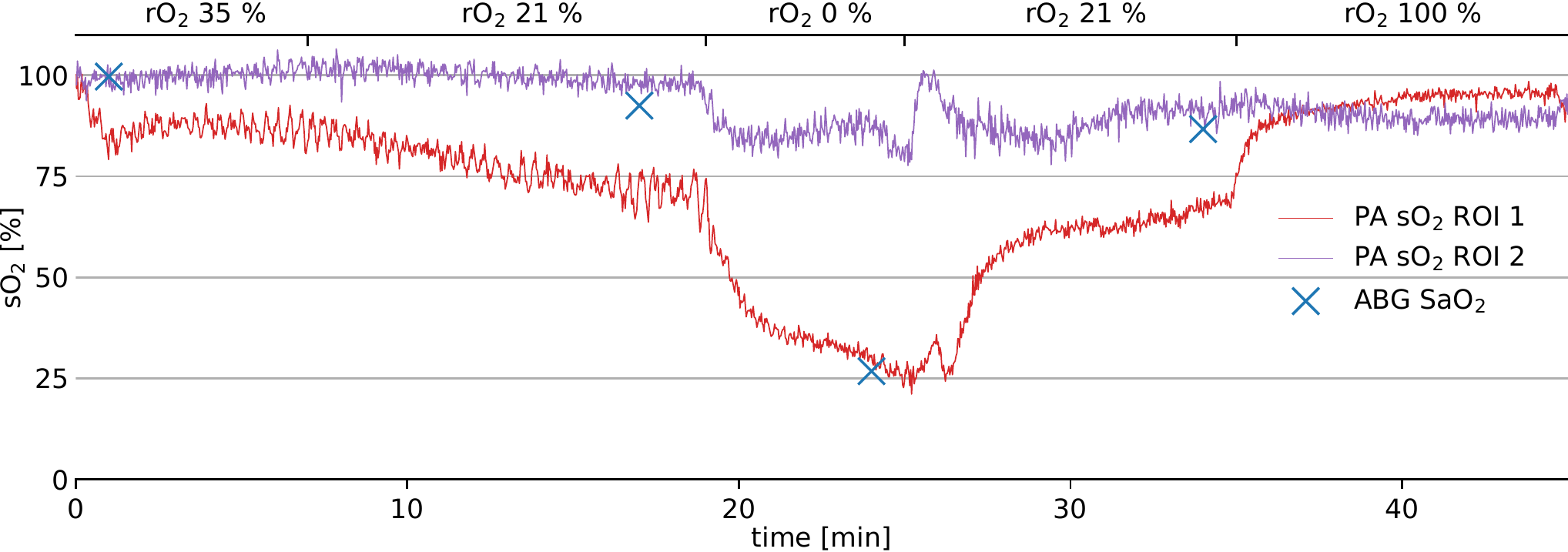}
\caption[]{Photoacoustic \acf{so2} estimation over time in two regions of interest (\acp{roi}) (see Figure \ref{fig:example_static_oxy}). The different levels of \acf{ro2} delivered by the ventilation are displayed above the plot; the \acf{abg} reference measurements of \acf{Sao2} as blue crosses.}
\label{fig:in_vivio}
\end{figure}

Comparing the changes in ventilation with the time course characteristics of the unmixing results of \ac{roi}\,1 in Figure \ref{fig:in_vivio}, one can see that linear \ac{su} can be used for change detection of \ac{so2} in the cerebral cortex. However, the unmixing results do not closely follow the quantitative values of the \ac{abg} reference (Table \ref{tab:ROI}). This illustrates the limits in quantification of \ac{so2} with \ac{pa} imaging and is even more obvious in deep tissue, considering the small changes in \ac{so2} estimation for \ac{roi}\,2.

\begin{table}[h!]
  \centering
    \footnotesize{\begin{tabular}{rrrrcrrrrrrrr}
    \toprule
    time & \ac{ro2} & \multicolumn{2}{c}{SaO$_2$\,[\%]} & & \multicolumn{2}{c}{QR/SVD sO$_2$\,[\%]} & \multicolumn{2}{c}{LU sO$_2$\,[\%]} & \multicolumn{2}{c}{NNLS sO$_2$\,[\%]} & \multicolumn{2}{c}{WLS sO$_2$\,[\%]} \\
    \cmidrule(l{2pt}r{2pt}){3-4} \cmidrule(l{2pt}r{2pt}){6-7} \cmidrule(l{2pt}r{2pt}){8-9} \cmidrule(l{2pt}r{2pt}){10-11} \cmidrule(l{2pt}r{2pt}){12-13}
    [min] & [\%] &\multicolumn{1}{l}{\acs{abg} } & \multicolumn{1}{c}{PuOx} & & \multicolumn{1}{c}{ROI 1} & \multicolumn{1}{c}{ROI 2} & \multicolumn{1}{c}{ROI 1} & \multicolumn{1}{c}{ROI 2} & \multicolumn{1}{c}{ROI 1} & \multicolumn{1}{c}{ROI 2} & \multicolumn{1}{c}{ROI 1} & \multicolumn{1}{c}{ROI 2} \\
    \midrule
    +0 & 35 & 100 & 99 & & 85 $\pm$ 2& 99 $\pm$ 2 & 85 $\pm$ 2 & 98 $\pm$ 2 & 85 $\pm$ 2 & 98 $\pm$ 2 & 85 $\pm$ 2 & 98 $\pm$ 2\\
    +8 & 21 & 93 & 88-92 & & 72 $\pm$ 3 & 98 $\pm$ 2 & 72 $\pm$ 4 & 98 $\pm$ 2 & 72 $\pm$ 3 & 98 $\pm$ 1& 72 $\pm$ 3 & 98 $\pm$ 2\\
   +24 & 0 & 26 & 40 & & 27 $\pm$ 2 & 84 $\pm$ 3 & 26 $\pm$ 2 & 83 $\pm$ 4 & 27 $\pm$ 2& 84 $\pm$ 3 & 27 $\pm$ 2& 84 $\pm$ 3\\
   +34 & 21 & 86 & -- & & 69 $\pm$ 2 & 91 $\pm$ 3 & 69 $\pm$ 2 & 91 $\pm$ 3 & 69 $\pm$ 2& 91 $\pm$ 3 & 69 $\pm$ 2& 91 $\pm$ 3\\
        +45 & 100 & -- & 100 & & 95 $\pm$ 2 & 91 $\pm$ 2 & 95 $\pm$ 2& 90 $\pm$ 2 & 95 $\pm$ 2 & 91 $\pm$ 2 & 95 $\pm$ 2& 90 $\pm$ 2\\
    \bottomrule
    \end{tabular}}%
    \caption{Comparison of the \acf{abg}, \acf{puox} and five \acf{su} estimations -- in \acf{roi} 1\,\&\,2. Values are averaged over one minute beginning at the time after start of the recording. The \acf{ro2} value was adjusted at least five minutes before the recording (see Figure\,2). SVD: singular value decomposition, NNLS: non-negative least squares, WLS: weighted least squares. --: Missing values failed to record due to technical issues.}
  \label{tab:ROI}%
\end{table}%

In conclusion, our study suggests that \ac{pa} imaging can be used to monitor \ac{so2} changes in the cerebral cortex during neurosurgical interventions. However, care must be taken when interpreting \ac{so2} estimation results due to the limits in quantitative accuracy when using linear \ac{su} algorithms. This is especially relevant in deep tissue due to fluence dependent spectral coloring. While there are promising approaches to address these fluence effects in general \cite{kirchner2018context, hanninen2018image, grohl2018confidence} and spectral coloring specifically \cite{tzoumas2016eigenspectra}, the translation of quantitative \ac{pa} imaging deep in tissue remains a major challenge.

%\paragraph{Abbreviations} --> *NO*
\begin{acronym}[MITK]
\acro{abg}[ABG]{arterial blood gas}
\acro{om}[OM]{Oxygenation measurements under hypoxic conditions}
\acro{roi}[ROI]{region of interest}
\acro{so2}[sO$_2$]{blood oxygenation}
\acro{Sao2}[SaO$_2$]{arterial blood oxygenation}
\acro{ro2}[rO$_2$]{respiratory oxygen}
\acro{su}[SU]{spectral unmixing}
\acro{pa}[PA]{photoacoustic}
\acro{mitk}[MITK]{Medical Interaction Toolkit}
\acro{us}[US]{ultrasonic}
\acro{puox}[PuOx]{pulse oximetry}
\end{acronym}

\section*{Acknowledgements} 
The authors would like to acknowledge support from the European Union through the ERC starting grant COMBIOSCOPY under the New Horizon Framework Programme grant agreement ERC-2015-StG-37960. The animal experiment was approved by the institutional animal care and use committee in Karlsruhe, Baden-W\"urttemberg, Germany; under Protocol No. 35-9185.81/G-174/16.

\begin{comment}
\paragraph{Contributions:} Conceptualization, T.K., J.G. and L.M.-H.; methodology, T.K.; software, T.K., J.G., N.H.;
data collection, T.K., J.G., M.A.H., A.H-A., E.S.; writing-original draft preparation, N.H.; writing-review and editing, T.K., J.G., N.H. and
L.M.-H.; supervision, L.M.-H.

\paragraph{Conflicts of Interest:} The authors declare no conflict of interest.
\end{comment}

\end{document}